\documentclass[12pt]{article}
\usepackage{latexsym}
\usepackage{amsfonts}
\usepackage{amssymb}
\usepackage[cp1250]{inputenc}

\title{A generalization of Margolus-Levitin bound}
\author{Bartosz Zieli\'nski  ,  
 Magdalena Zych\\  
Department of Theoretical Physics II,\\Institute of Physics,
University of {\L}\'od\'z, \\
ul. Pomorska 149/153, 90 - 236 {\L}\'od\'z, Poland.}
\date{}

\begin{document}
\maketitle
\begin{abstract}
The Margolus-Levitin lower bound on minimal time required for a state to be transformed into an orthogonal state 
is generalized. It is shown that for some initial 
states new bound is stronger than the Margolus-Levitin one.
\end{abstract}
\newpage
A useful measure of the evolution speed of quantum systems is the minimal time $t_1$\ required for a state to be transformed 
into an orthogonal state. There exist two basic estimates of $t_1$. 

First, $t_1$\ obeys
\begin{eqnarray}
t_1\geq \frac{\pi \hbar}{2\Delta E}  \label{w1}
\end{eqnarray}
where $\Delta E$\ is energy dispersion of initial state. Eq. (\ref{w1}) follows easily from the inequality derived by 
Mandelstam and Tamm and was studied by many authors $\cite {b2}\div \cite{b7}$. 

The second estimate has been derived few years ago by Margolus and Levitin \cite{b8}. It is valid for Hamiltonians 
bounded from below and reads
\begin{eqnarray}
t_1\geq \frac{\pi \hbar}{2\langle E-E_0\rangle } ;  \label{w2}
\end{eqnarray}
here $\langle \;\;\; \rangle $\ denotes the initial state expectation value while $E_0$\ is a ground - state energy.
 
Both eqs. (\ref{w1}) 
and (\ref{w2}) can be derived using similar arguments \cite{b8}, \cite{b9}. By virtue of the spectral theorem one writes
\begin{eqnarray}
&& \langle \Psi \mid e^{\frac{-itH}{\hbar}}\mid \Psi \rangle =\int e^{\frac{-itE}{\hbar}}d \langle \Psi \mid 
P_E\mid \Psi \rangle = \nonumber \\
&& =\int \cos\left(\frac{tE}{\hbar}\right)d \langle \Psi \mid P_E\mid \Psi \rangle -
i\int \sin\left(\frac{tE}{\hbar}\right)d \langle \Psi 
\mid P_E\mid \Psi \rangle  \label{w3}
\end{eqnarray}
where $P_E$\ is the spectral measure which enters spectral decomposition of $H$, $H=\int EdP_E$.
Therefore, denoting $ \langle \Psi \mid A\mid \Psi \rangle \equiv \langle A \rangle $, one gets 
\begin{eqnarray}
\left\langle \cos\left(\frac{t_1H}{\hbar}\right)\right \rangle =0= \left\langle \sin\left(\frac{t_1H}{\hbar}\right)\right \rangle   \label{w4}
\end{eqnarray}
Now, consider an inequality of the form
\begin{eqnarray}
f(x)\geq A \sin x+B \cos x   \label{w5}
\end{eqnarray}
which is assumed to hold for all $x\geq 0$ (actually, in order to prove (\ref{w1}) one demands (\ref{w5})
 to hold for all $x$).
 Denoting by $E_0$\ the lower energy bound one finds from eq. (\ref{w5})
\begin{eqnarray}
\left\langle f\left(\frac{t(H-E_0)}{\hbar}\right)\right \rangle \geq
 A \left\langle \sin \left(\frac{t(H-E_0)}{\hbar}\right) \right\rangle 
+B \left\langle  
\cos \left(\frac{t(H-E_0)}{\hbar}\right) \right\rangle   \label{w6}
  \end{eqnarray}
provided $\mid \Psi \rangle $\ belongs to the domain of $f\left(\frac{t(H-E_0)}{\hbar}\right)$.
Indeed, eq.(\ref{w6}) follows easily from the inequality (\ref{w5}) by noting that the expectation value of nonnegative
 function is nonnegative
\begin{eqnarray}
&&\left\langle f\left(\frac{t(H-E_0)}{\hbar}\right) - A\sin\left(\frac{t(H-E_0)}{\hbar}\right) - B\cos\left(\frac{t(H-E_0)}{\hbar}\right)\right\rangle =\nonumber \\
&&= \int \Big[f\left(\frac{t(E-E_0)}{\hbar}\right) - A\sin\left(\frac{t(E-E_0)}{\hbar}\right) - \nonumber \\
&& -B\cos\left(\frac{t(E-E_0)}{\hbar}\right)\Big]
d\langle\Psi \mid P_E\mid \Psi\rangle \geq  0  \label{w17}
\end{eqnarray} 
 In particular, eqs. (\ref{w4}) 
and (\ref{w6})  imply
\begin{eqnarray}
\left\langle f\left(\frac{t_1(H-E_0)}{\hbar}\right) \right\rangle \geq 0  \label{w7}
\end{eqnarray}
which imposes some restrictions on $t_1$. \\

In order to derive a new bound on $t_1$\ we use the following inequality
\begin{eqnarray}
x^{\alpha }-\frac{\pi ^{\alpha }}{2}+\frac{\pi ^{\alpha }}{2}\cos x+\alpha \pi ^{\alpha -1}\sin x \geq 0   \label{w8}
\end{eqnarray}
which holds for all $x\geq 0$\ and $\alpha >0$. Note that eq.(\ref{w8}) provides a generalization of the inequality used 
in Ref. \cite{b8}; it reduces to the latter for $\alpha =1$. \\
By virtue of eq. (\ref{w7}), eq. (\ref{w8}) leads to the following bound on $t_1$:
\begin{eqnarray}
t_1\geq \frac{\pi \hbar}{2^{\frac{1}{\alpha }}\langle (E-E_0)^{\alpha } \rangle^{\frac{1}{\alpha }}}, \;\;\; \alpha >0  \label{w10}
\end{eqnarray}
provided $\mid \Psi  \rangle $\ belongs to the domain of $(H-E_0)^{\alpha }$. Eq. (\ref{w10}) provides the generalization 
of Margolus - Levitin bound which is attained for $\alpha =1$. 

The estimate (\ref{w10}) is for fixed $\alpha \neq 1$\ neither weaker nor stronger than the Margolus - Levitin one. Indeed, 
although the convexity (concavity) of $x\rightarrow x^{\alpha }$\ for $\alpha >1 \;\;\; (\alpha <1)$\ allows us to claim 
that $\langle E^{\alpha } \rangle ^{\frac{1}{\alpha }}\geq  \langle E \rangle \;\;\; ( \langle E^{\alpha } \rangle ^{\frac{1}
{\alpha }} \leq  \langle E \rangle ) $, the additional factor $2^{\frac{1}{\alpha }}$\ makes  apriori estimate impossible. 
Obviously, one could take the supremum over all $\alpha >0$\ of the right hand side of (\ref{w10}). However, this is only 
possible for $\mid \Psi  \rangle $\ belonging to the domains of all $(H-E_0)^{\alpha }, \;\; \alpha >0 $. \\

In order to show that, in some cases, the inequality (\ref{w10}) gives much better bound for some $\alpha \neq 1$\ one can 
use a simple example considered in Ref. \cite{b8}. Let us take the initial state of the form
\begin{eqnarray}
\mid \Psi \rangle =\frac{a}{\sqrt{2}}(\mid 0\rangle +\mid \varepsilon \rangle )+\frac{b}{\sqrt{2}}(\mid n\varepsilon \rangle +
\mid (n+1)\varepsilon \rangle );   \label{w11}
\end{eqnarray}
normalization condition implies $\mid a\mid ^2+\mid b\mid ^2=1$. \\
One easily checks that
\begin{eqnarray}
t_1=\frac{\pi \hbar}{\varepsilon }  \label{w12}
\end{eqnarray}
Computing the relevant expectation value one obtains
\begin{eqnarray}
2^{\frac{1}{\alpha }}\langle (E-E_0)^{\alpha } \rangle^{\frac{1}{\alpha }}
 =(1+\mid b\mid ^2(n^{\alpha }+(n+1)^{\alpha }-1))^{\frac{1}{\alpha }}
\varepsilon   \label{w13}
\end{eqnarray}
Let us choose $b=\lambda / \sqrt{2}\sqrt[4]{n}$\ with $\lambda \neq 0$\ independent of $n$. Then, for $\alpha =\frac{1}{2}$,  
eq.(\ref{w10}) gives in the limit of large $n\;(\sqrt{n}\gg 1)$\
\begin{eqnarray}
t_1\geq \frac{\pi \hbar}{\varepsilon (1+\mid \lambda \mid ^2)^2}  \label{w14}
\end{eqnarray}
On the other hand, if  $\mid \lambda \mid ^2\sqrt{n}\gg 1$, eq.(\ref{w2}) becomes
\begin{eqnarray}
t_1\geq \frac{\pi \hbar}{\varepsilon \mid \lambda \mid ^2\sqrt{n}}  \label{w15}
\end{eqnarray}
Also eq.(\ref{w1}) gives in this limit much weaker bound
\begin{eqnarray}
t_1\geq \frac{\sqrt{2}\pi \hbar}{\varepsilon \mid \lambda \mid \sqrt[4]{n^3}}     \label{w16}
\end{eqnarray}
We see that for the above state our bound is $O(1)$\ while (\ref{w1}) and (\ref{w2}) are $O(\frac{1}{\sqrt[4]{n^3}})$\
and $O(\frac{1}{\sqrt{n}})$, respectively.  Therefore, the new bound may be much better even for such very simple
 systems. 
 
 The above example may seem quite artificial. However, it is generic in the sense that
 it allows us to understand the status of bounds based on
energy distribution moments. In fact, let us consider the following generalization of our example. We assume that 
the energy spectrum consists of a number of pairs of levels differing by the same energy amount $\varepsilon $:
$spec(H) = \{ E_0=0, \varepsilon , E_1, E_1+\varepsilon , E_2, E_2+\varepsilon , ...\}$. Consider the state for which
both members of any "doublet" enter with the same amplitude, i.e. 
\begin{eqnarray} 
\mid \Psi \rangle = \sum_n \frac{a_n}{\sqrt{2}}(\mid E_n \rangle + \mid E_n+\varepsilon  \rangle),\;\;\;\;
\sum_n \mid a_n \mid ^2 = 1    \label{w18}
\end{eqnarray}
 Obviously, the orthogonalization time for this state is given by eq.(\ref{w12}), irrespectively of the values of
$a_n$\ and $E_n$, $n=0,1,2,...$. On the other hand
\begin{eqnarray}
 2^{\frac{1}{\alpha }}\langle (E-E_0)^{\alpha } \rangle^{\frac{1}{\alpha }} =
 \left(\sum_n \mid a_n \mid ^2(E_n^\alpha  + (E_n+\varepsilon )^\alpha )\right)^{\frac{1}{\alpha }}    \label{w19}
\end{eqnarray}
It is clearly seen from the above equation that our bound cannot be optimal except for the small number 
of states (see below). However, the advantage of it is that we have a free parameter $\alpha $\ which can be manipulated
to get the best possible estimate for known spectrum. As we have shown explicitly above an appropriate choice of 
$\alpha $\ can result in much better bound than Margolus-Levitin one.

The above reasoning shows also clearly that there exists no optimal bound based on energy distribution only.
The relevant moments generically depend strongly of the values $E_k$\ and $a_k$\ which, in turn, are completely irrelevant 
as far as the orthogonalization time is concerned. Therefore, it is desirable to have an apriori estimates which 
depend on free parameter to be adjusted to "minimalize" the role of $E_k$\ and $a_k$.

Let us find the intelligent states saturating (\ref{w10}). To this end let us note that the LHS of eq.(\ref{w8}) 
vanishes only for $x=0$\ and $x=\pi $. Therefore, only two - level systems can saturate (\ref{w10}). One easily finds that 
they must be of the form
\begin{eqnarray}
\mid \Psi \rangle =c_1\mid E_0 \rangle +c_2\mid E_1\rangle ,\;\;\; \mid c_1\mid =\mid c_2\mid =\frac{1}{\sqrt{2}}  \label{w17}
\end{eqnarray}
Finally, let us sketch how one can generalize our result to mixed states case. This can be done according to the lines of 
ref. \cite{b10}. To this end, given two density matrices,  one defines the fidelity 
\begin{eqnarray}
F(\rho ,\rho ') = \left(Tr \sqrt{\sqrt{\rho }\rho '\sqrt{\rho }}\right)^2  \label{w20}
\end{eqnarray}
Given any Hamiltonian $H$\ and an initial state $\rho $\ 
\begin{eqnarray}
\rho = \sum_np_n\mid \phi _n\rangle \langle \phi _n\mid    \label{w22}
\end{eqnarray}

we want to estimate the value of $F(\rho , \rho (t))$.
To this end we consider some purification $\mid \chi \rangle$\ of $\rho $,
\begin{eqnarray}
\mid \chi \rangle = \sum_n\sqrt{p_n} \mid \phi _n \rangle \mid \xi _n \rangle .     \label{w23}
\end{eqnarray}
 Assume that all states of an ancillary 
system evolve trivially in time. Then the total Hamiltonian governing the time evolution of $\mid \chi \rangle$\ 
equals $H\otimes I$. Therefore, all energy distribution moments with respect to $\mid \chi \rangle$\ coincide with 
those with respect to $\rho $. Due to the Uhlmann's theorem \cite{b11} the following inequality holds 
\begin{eqnarray}
F(\rho ,\rho (t)) \geq \mid \langle\chi \mid \chi (t) \rangle\mid ^2    \label{w21}
\end{eqnarray}

which allows us to  extend to the mixed state case any bound based on energy distribution moments.

\vspace {12pt}
{\large\bf Acknowledgement}

This research was supported by the University of {\L}\'od\'z grants $N^o$\ $690$\  and $795$.
We thank the unknown referee for bringing ref.[10] to our attention.


\begin{thebibliography}{99}
\bibitem{b1}
L. Mandelstam, I. Tamm, Journ. Phys. (USSR) {\bf 9} (1945), 249
\bibitem{b2}
G.N. Fleming, Nuovo Cim. {\bf A16} (1973), 232
\bibitem{b3}
K. Bhattacharyya, Journ. Phys. {\bf A16} (1983), 2991
\bibitem{b4}
D. Home, M.A.B. Whitaker, Journ. Phys. {\bf A19} (1986), 1847
\bibitem{b5}
L. Vaidman, Am. Journ. Phys. {\bf 60} (1992), 182
\bibitem{b6}
L. Vaidman, O. Belkind, Phys. Rev. {\bf A57} (1998), 1583
\bibitem{b7}
A. Peres, Quantum Theory: Concepts and Methods, Kluwer, Hingham 1985
\bibitem{b8}
N. Margolus, L.B. Levitin, Physica {\bf D120} (1998), 188
\bibitem{b9}
P. Kosi\' nski, M. Zych,  Phys. Rev. {\bf A73} (2006), 024303
\bibitem{b10}
V. Giovannetti, S. Lloyd, L. Maccone, Phys. Rev. {\bf A67} (2003), 052109
\bibitem{b11}
A. Uhlmann, Rep. Math. Phys. {\bf 9} (1976), 273  
%
%








\end{thebibliography}
\end{document}